\documentclass[12pt]{iopart}
\usepackage[numbers]{natbib}

\usepackage{iopams}

\usepackage{xcolor}
\usepackage{graphicx}
\usepackage[colorlinks]{hyperref}
\usepackage{caption}
\usepackage{subcaption}

\newcommand{\revise}[1]{\textcolor{black}{#1}}
\newcommand{\newrevise}[1]{\textcolor{black}{#1}}

\begin{document}

\title[Enhancing a CAE with a QAOA for Image Noise Reduction]{Enhancing a Convolutional Autoencoder with a Quantum Approximate Optimization Algorithm for Image Noise Reduction}

\author{Tara Kit$^1$, Kimsay Pov$^1$, Kimleang Kea$^1$, Won-Du Chang$^1$, Hee Chul Park$^2$, and Youngsun Han$^{1,*}$}

\address{$^1$ Department of AI Convergence, Pukyong National University, Busan 48513, South Korea}
\address{$^2$ Department of Physics, Pukyong National University, Busan 48513, South Korea}
\address{$^*$ Authors to whom any correspondence should be addressed.}

\ead{youngsun@pknu.ac.kr}

\begin{abstract}
Image denoising is essential for removing noise in images caused by electric device malfunctions or other factors during image acquisition. It ensures the preservation of image quality and accurate interpretation. Many convolutional autoencoder algorithms have proven effective in image denoising. Owing to their promising efficiency, quantum computers have gained popularity. \revise{This paper proposes a method, the quantum convolutional autoencoder (QCAE), which enhances traditional convolutional autoencoders by replacing their latent space with a quantum counterpart implemented via a QAOA-inspired ansatz circuit.} To enhance efficiency, we leveraged the advantages of the quantum approximate optimization algorithm (QAOA)-incorporated parameter-shift rule to identify an optimized cost function, facilitating effective learning from data and gradient computation on an actual quantum computer. The proposed QCAE method outperformed its classical counterpart as it exhibited lower training loss and a higher structural similarity index (SSIM) value. QCAE also outperformed its classical counterpart in denoising the MNIST dataset by up to 40\% in terms of SSIM value, confirming its enhanced capabilities in real-world applications. Evaluation of QAOA performance across different circuit configurations and layer variations showed that our technique outperformed other circuit designs by 25\% on average.
\end{abstract}

\vspace{2pc}

\submitto{\textit{Mach. Learn.: Sci.}}

\maketitle

\section{Introduction}\label{sec:introduction}
Images acquired from image sensors frequently exhibit noise caused by hardware malfunctions and various environmental factors. In this context, noise denotes unpredictable fluctuations in brightness, intensity, or color information within a digital image~\cite{diwakar2018review, paul2022modified}. Image denoising is pivotal for eliminating these distortions from a given input image, aiming to generate a more refined and clearer image~\cite{elad2023image, izadi2023image}. The denoising process is essential for accurate image analysis and benefits both human interpretation and machine learning (ML)-based assessments~\cite{wang2022adaptive}.

\revise{An autoencoder (AE) is an unsupervised neural network model designed for reconstructing its input through a learned latent representation.}
This class of models demonstrates notable versatility, particularly in the group of learning representations~\cite{abiodun2018state}. AEs encompass numerous successful variants, including the variational autoencoder (VAE), convolutional autoencoder (CAE), and denoising autoencoder, as discussed in~\cite{sagha2017stacked}. They operate by mapping the input to a lower-dimensional space (latent space). These encoding and decoding steps are subsequently used to accurately reconstruct the original image. However, the latent space in AE encounters challenges in accurately capturing complex, nonlinear relationships among data points. \revise{Moreover, it often prioritizes frequent data patterns, resulting in sparse or poorly structured latent representations that underrepresent rare samples. This can reduce effectiveness in generative or reconstruction tasks~\cite{Leeb2021ExploringTL}.}

Recently, the popularity of quantum computers has surged among researchers, driven by their unique computational capabilities, such as superposition. 
\revise{Quantum circuits leverage principles such as superposition, entanglement, and interference, allowing them to represent and manipulate complex, high-dimensional states that are often intractable for classical systems.}
This capability can lead to more accurate and efficient data compression than classical AE. However,  quantum computers remain suboptimal for production applications~\cite{bouland2022noise}. The existing quantum computing (QC) landscape is in the noisy intermediate-scale quantum (NISQ) era~\cite{bharti2022noisy}, indicating that current quantum systems exhibit limitations in terms of the number of qubits, circuit depth, and high noise levels. Because of these factors, variational quantum algorithms (VQAs) have emerged as an ideal choice for implementation, primarily due to their lower qubit requirements and reduced circuit depth. VQAs leverage classical computation to optimize the algorithm parameters effectively~\cite{huang2023near}. In addition, they exhibit robustness against image noise - a major problem for other types of quantum algorithms. 
\revise{We adopt a QAOA-inspired ansatz circuit, leveraging its parameterized structure to construct a quantum latent space representation, rather than applying QAOA for solving an optimization task.}
\revise{These quantum principles, superposition, entanglement, and interference, allow the circuit to represent and process complex, high-dimensional patterns in data. In our approach, this variational structure enables effective learning and denoising of noisy images through gradient-based optimization, rather than through solving a combinatorial objective.}

Due to exponentially increasing computational power and the ability to learn data representations in higher-dimensional spaces, our primary objective in this study is to integrate CAE and QAOA to create an advanced latent space for execution within QC. We enhanced our model by incorporating a trainable quantum variational circuit to develop a hybrid architecture termed a quantum convolutional autoencoder (QCAE). The QCAE was specifically designed for the task of MNIST~\cite{mnistdata} image denoising. To showcase its efficacy, we conducted a comparative analysis of its performance against its classical counterpart, the classical convolutional autoencoder (CCAE). The contributions of this study are as follows:
\begin{itemize}
    \item \newrevise{We present an approach where the convolutional autoencoder ensures efficient feature extraction, while the QAOA-inspired ansatz circuit contributes the advantage of reduced qubit requirements and improved suitability for NISQ devices.}
    \item \revise{We leverage a QAOA-inspired ansatz circuit to replace the classical latent space, enabling the representation of input data in a higher-dimensional quantum space for improved feature extraction and denoising.}
    \item \revise{We apply the parameter-shift rule (PSR) to efficiently and precisely compute gradients of the QAOA-inspired ansatz circuit, enabling stable optimization within hybrid quantum-classical training.}
    \item We rigorously conducted denoising experiments using the MNIST dataset, exploring diverse image noise \revise{(Gaussian noise)} factors across both quantum simulation and machine scenarios.
\end{itemize}

The rest of the paper is organized as follows. Section~\ref{sec:related_works} describes both classical and quantum algorithms for image denoising and classification. Section~\ref{sec:preliminaries} reviews the associated background to understand this paper. In Section~\ref{sec:method}, we delve into the details of our proposed methodology, leveraging QAOA in conjunction with PSR optimization. Section~\ref{sec:setup} details the specifics of the dataset and environments employed in our study. In Section~\ref{sec:results}, we present a comprehensive showcase of our superior performance in image denoising. Finally, Section~\ref{sec:conclusion} concludes the paper.

\section{Related works}\label{sec:related_works}
In the field of classical image denoising, deep learning serves as the main building block, harnessing the power of neural networks (NN) to effectively reduce noise and enhance image quality.

\newrevise{In the field of classical image denoising, deep learning serves as the main building block, harnessing the power of neural networks to effectively reduce noise and enhance image quality~\cite{elad2023image, izadi2023image}}. Autoencoders and their variants, including convolutional autoencoders, have been widely applied to natural images and benchmark datasets, demonstrating strong denoising performance~\cite{schneider2022autoencoders}. These methods form the baseline context against which hybrid quantum–classical approaches, such as our QCAE, can be evaluated.
% Paul et al.~\cite{paul2022wavelet} introduced an efficient denoising method for hyperspectral images (HSI), targeting various noise patterns and their combinations. Using a dual-branch deep neural network based on wavelet-transformed bands, the first branch uses convolutional layers with residual learning for local and global noise feature extraction. The second branch contains an AE layer with subpixel sampling for prominent noise feature extraction. Schneider et al.~\cite{schneider2022autoencoders} implemented a CAE, a VAE, and an adversarial autoencoder on two distinct publicly available datasets.

Numerous studies leverage QC and NN to develop image classification and denoising. Bravo et al.~\cite{bravo2021quantum} presented an enhanced feature quantum autoencoder, a VQA capable of compressing the quantum states of different models with higher fidelity. They validated the method in simulations by compressing the ground states of the Ising model and classical MNIST handwritten digits. Although this approach effectively compressed and reconstructed images, its primary design did not prioritize the task of denoising images. Rivas et al.~\cite{rivas2021hybrid} presented a hybrid QML approach for representation learning. They utilized a quantum variational circuit that could be trained using conventional gradient descent techniques. To further enhance its performance, the method dressed the quantum circuit with AE layers for meaningful interpretation of complex unstructured data. Orduz et al.~\cite{orduz2021quantum} applied randomized quantum circuits to process input data, functioning as quantum convolutions to generate representations applicable in convolutional networks. The experimental findings indicate comparable performance to traditional convolutional neural networks, with quantum convolutions demonstrating potential acceleration in convergence in certain scenarios. Nguyen et al.~\cite{nguyen2018image} used a D-Wave 2X quantum computer to address NP-hard sparse coding problems in image classification. The method involves downsampling MNIST images, employing a bottleneck autoencoder, and using quantum inference on D-Wave 2X to obtain sparse binary representations. The study also explored lateral inhibition between features generated from a subset of autoencoder-reduced MNIST images. Sleeman et al.~\cite{sleeman2020hybrid} proposed a hybrid system that integrates a classical AE with a quantum annealing restricted Boltzmann machine (RBM) on a D-Wave quantum computer. The system achieved a nearly 22-fold compression of grayscale images to binary with lossy recovery. The evaluation of the MNIST dataset demonstrated the feasibility of using D-Wave as a sampler for ML. Additionally, the study compared the results of a hybrid quantum RBM with a hybrid classical RBM in a downstream classification problem and measured image similarity between MNIST-generated images and the original training dataset.

\section{Preliminaries}\label{sec:preliminaries}
In this section, we introduce the basics of QC and its data processing. We then explore how quantum computers handle classical data, discuss the execution of parameterized circuit data, and present the foundation of CCAE as our proposed method counterpart.

\subsection{Quantum computing (QC)}
\revise{Quantum computing leverages fundamental principles such as superposition, entanglement, and quantum interference to explore new computational paradigms, with the potential for advantages in specific tasks over classical systems}
~\cite{national2019quantum}.

\subsubsection{Quantum bit (Qubit)}
The operational architecture of quantum computers differs fundamentally from that of classical computers. Quantum computers use quantum bits (qubits) as operational units, unlike classical computers, which rely on binary digits existing in either state 0 or 1. A qubit can exist simultaneously in a superposition of states 0, 1, and a linear combination of both 0 and 1~\cite{nielsen2010quantum}.
\revise{The basis states of $\vert 0 \rangle$ and $\vert 1 \rangle$ represent the computational basis states of a qubit, forming an orthonormal basis in a two-dimensional Hilbert space. Any pure qubit state can be expressed as a linear combination of these basis states, defined as follows:}
% \begin{equation}
%     \vert 0 \rangle = 
%     \begin{bmatrix}
%                 1\\
%                 0
%     \end{bmatrix},\;
%     \vert 1 \rangle = 
%     \begin{bmatrix}
%                 0\\
%                 1
%     \end{bmatrix},\;
%     \vert \psi \rangle = \alpha \vert 0 \rangle + \beta \vert 1 \rangle =
%     \begin{bmatrix}
%                 \alpha \\
%                 \beta
%     \end{bmatrix},
% \end{equation}
\begin{equation}
\left|0\right\rangle =
\left[\!
\begin{array}{c}
1\\[2pt]
0
\end{array}
\!\right],
\quad
\left|1\right\rangle =
\left[\!
\begin{array}{c}
0\\[2pt]
1
\end{array}
\!\right],
\quad
\left|\psi\right\rangle = \alpha\left|0\right\rangle + \beta\left|1\right\rangle =
\left[\!
\begin{array}{c}
\alpha\\[2pt]
\beta
\end{array}
\!\right].
\end{equation}

where the probability coefficients $\alpha$ and $\beta$ are called \textit{amplitudes}. The sum of the squared amplitudes in a superposition equals 1, which is ${\vert \alpha \vert}^2 + {\vert \beta \vert}^2 = 1$. Despite the ability of qubits to exist simultaneously in both $\vert 0 \rangle$ and $\vert 1 \rangle$ states, their definitive output is determined as either $\vert 0 \rangle$ or $\vert 1 \rangle$ when measured.

\subsubsection{Quantum data manipulation}
Data on a qubit can be transformed and manipulated using quantum gates. These quantum gates fundamentally involve the modification of one or more qubits~\cite{stein2021hybrid}. The gates introduced in this paper include Hardamard ($H$), CNOT, and rotation gates such as X-rotation ($R_X$), Y-rotation ($R_Y$), and Z-rotation ($R_Z$). They are defined as follows:
% \begin{equation}
% \begin{array}{ll}
%     H = \frac{1}{\sqrt{2}}
%     \begin{bmatrix}
%                 1 & 1\\
%                 1 & -1
%     \end{bmatrix}, \quad
%     CNOT =
%     \begin{bmatrix}
%                 1 & 0 & 0 & 0\\
%                 0 & 1 & 0 & 0\\
%                 0 & 0 & 0 & 1\\
%                 0 & 0 & 1 & 0\\
%     \end{bmatrix},
% \end{array}
% \end{equation}
% \begin{equation}
%     R_X(\theta) =
%     \begin{bmatrix}
%                 cos\Bigl(\frac{\theta}{2}\Bigl) & -i \cdot sin\Bigl(\frac{\theta}{2}\Bigl) \\ \\
%                 -i \cdot sin\Bigl(\frac{\theta}{2}\Bigl) & cos\Bigl(\frac{\theta}{2}\Bigl)
%     \end{bmatrix},
% \end{equation}
% \begin{equation}
%     R_Y(\theta) =
%     \begin{bmatrix}
%                 cos\Bigl(\frac{\theta}{2}\Bigl) & -sin\Bigl(\frac{\theta}{2}\Bigl) \\ \\
%                 sin\Bigl(\frac{\theta}{2}\Bigl) & cos\Bigl(\frac{\theta}{2}\Bigl)
%     \end{bmatrix},
% \end{equation}
% \begin{equation}
%     R_Z(\theta) =
%     \begin{bmatrix}
%                 e^{-i\frac{\theta}{2}} & 0 \\ 
%                 0 & e^{i\frac{\theta}{2}}
%     \end{bmatrix}.
% \end{equation}
\begin{equation}
\begin{array}{ll}
H = \frac{1}{\sqrt{2}}
\left[\!
\begin{array}{cc}
1 & 1\\[2pt]
1 & -1
\end{array}
\!\right],
\quad
\mathrm{CNOT} =
\left[\!
\begin{array}{cccc}
1 & 0 & 0 & 0\\[2pt]
0 & 1 & 0 & 0\\[2pt]
0 & 0 & 0 & 1\\[2pt]
0 & 0 & 1 & 0
\end{array}
\!\right].
\end{array}
\end{equation}

\begin{equation}
R_X(\theta) =
\left[\!
\begin{array}{cc}
\cos\!\Bigl(\frac{\theta}{2}\Bigr) & -i\,\sin\!\Bigl(\frac{\theta}{2}\Bigr)\\[4pt]
-i\,\sin\!\Bigl(\frac{\theta}{2}\Bigr) & \cos\!\Bigl(\frac{\theta}{2}\Bigr)
\end{array}
\!\right].
\end{equation}

\begin{equation}
R_Y(\theta) =
\left[\!
\begin{array}{cc}
\cos\!\Bigl(\frac{\theta}{2}\Bigr) & -\sin\!\Bigl(\frac{\theta}{2}\Bigr)\\[4pt]
\sin\!\Bigl(\frac{\theta}{2}\Bigr) & \cos\!\Bigl(\frac{\theta}{2}\Bigr)
\end{array}
\!\right].
\end{equation}

\begin{equation}
R_Z(\theta) =
\left[\!
\begin{array}{cc}
e^{-i\theta/2} & 0\\[4pt]
0 & e^{i\theta/2}
\end{array}
\!\right].
\end{equation}

All gates except CNOT are single-qubit gates. The $H$ gate enables the creation of a superposition for a single qubit, while rotation gates allow a qubit to be positioned at any point on a Bloch sphere surface. However, CNOT is a two-qubit gate that facilitates entanglement between two qubits. These gates collectively enable quantum state manipulation.

\subsection{Classical data encoding}
\begin{figure}[ht]
    \centering
    \includegraphics[width=0.3\textwidth]{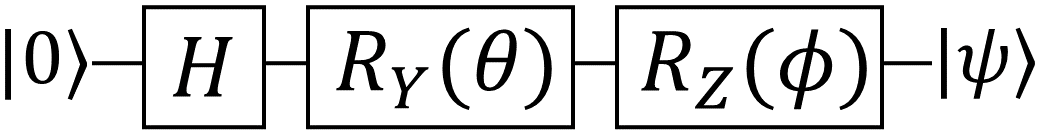}
    \caption{Angle encoding quantum circuit.}
    \label{fig:angle_encoding}
\end{figure}

Numerous QML algorithms require converting classical data into quantum states within quantum computers, a process known as data encoding~\cite{cortese2018loading, wang2022development}. In other words, the dataset is initially translated from the subject data domain $D$ to the Hilbert space $H$ through a designated feature mapping process $f: D \rightarrow H$. One of the encoding techniques, specifically angle encoding or qubit encoding, is widely renowned for its ability to leverage a minimal number of qubits that correspond to the size of the input vector~\cite{ovalle2023quantum}. The angle encoding scheme demonstrates remarkable efficiency as it necessitates the rotation of only a single qubit, as depicted in Fig.~\ref{fig:angle_encoding}. 

In this technique, each data value $x$ undergoes an initial normalization process, mapping it to the range $[0,2\pi]$. Subsequently, it is encoded using single-qubit rotation gates $R_X$, $R_Y$, and $R_Z$. These rotation gates dynamically determine the rotation angle $\theta$ based on the corresponding data value $x$. The predominant technique involves applying the feature map $x \rightarrow cos\Bigl(\frac{x}{2}\Bigl) \vert 0 \rangle + sin\Bigl(\frac{x}{2}\Bigl) \vert 1 \rangle$ using the $R_Y$ as follows:
% \begin{equation}
% \begin{array}{ll}
%     R_Y(x)\vert 0 \rangle &=
%     \begin{bmatrix}
%                 cos\Bigl(\frac{x}{2}\Bigl) & -sin\Bigl(\frac{x}{2}\Bigl) \\ \\
%                 sin\Bigl(\frac{x}{2}\Bigl) & cos\Bigl(\frac{x}{2}\Bigl)
%     \end{bmatrix}
%     \begin{bmatrix}
%                 1\\ \\
%                 0
%     \end{bmatrix} \\ \\
%     &= cos\Bigl(\frac{x}{2}\Bigl) \vert 0 \rangle + sin\Bigl(\frac{x}{2}\Bigl) \vert 1 \rangle .
% \end{array}
% \end{equation}
\begin{equation}
\begin{array}{ll}
R_Y(x)\left|0\right\rangle &=
\left[\!
\begin{array}{cc}
\cos\!\Bigl(\frac{x}{2}\Bigr) & -\sin\!\Bigl(\frac{x}{2}\Bigr)\\[4pt]
\sin\!\Bigl(\frac{x}{2}\Bigr) & \cos\!\Bigl(\frac{x}{2}\Bigr)
\end{array}
\!\right]
\left[\!
\begin{array}{c}
1\\[4pt]
0
\end{array}
\!\right]\\[10pt]
&= \cos\!\Bigl(\frac{x}{2}\Bigr)\left|0\right\rangle
   + \sin\!\Bigl(\frac{x}{2}\Bigr)\left|1\right\rangle.
\end{array}
\end{equation}

As such, this approach requires $n$ qubits to encode $n$ input variables defined as follows:
\begin{equation}
    \vert \psi_{x} \rangle = \bigotimes_{i=1}^{n} R(x_i) \vert \psi_{0} \rangle ,
\end{equation}
where $R$ is one of the rotation gates and $\psi_{0}$ is an initial state.

\subsection{Parameterized quantum circuit (PQC)}
% A PQC provides a tangible means to implement algorithms and showcase quantum supremacy in the NISQ era
\revise{A PQC provides a practical framework for implementing variational algorithms on NISQ devices, enabling hybrid quantum-classical computations with limited qubits and circuit depth}~\cite{benedetti2019parameterized, hubregtsen2021evaluation}. Conceptually similar to a classical NN, it features trainable parameters within circuits comprising fixed gates, such as CNOT, and adjustable gates like rotation gates. It typically requires a low number of qubits and circuit depth. However, even with a limited number of qubits and circuit depth, certain classes of PQCs exhibit the ability to generate highly non-trivial outputs. Similar to classical NN, a higher circuit depth repeated in a PQC provides increased flexibility to the circuit, leading to superior results in various applications.

Consider a trainable unitary $U_{\theta}$ operating on an $n$-qubit state applied to a reference state $\vert \phi \rangle$, where the trained variable $\vert \phi _{\theta} \rangle = U_{\theta}\vert \phi \rangle$ generally equals to $\vert 0 \rangle^{\otimes n}$. Nevertheless, in the context of an $n$-qubit encoding, which represents a $2^n \times 2^n$ unitary mapping, the number of input qubits equals the number of output qubits. For the 8-qubit circuit, we encounter a total of $2^{16} = 65,536$ parameters, which proves to be an excessively high number for efficient optimization. \revise{Therefore, to enhance efficiency, we constrained gate shapes to have fewer parameters using small, efficient gates such as single-qubit rotations and entangling gates like CNOT as the fundamental building blocks of the overall unitary operation.}

\subsection{Classical convolutional autoencoder (CCAE)}
\begin{figure}[ht]
    \centering
    \includegraphics[width=0.47\textwidth]{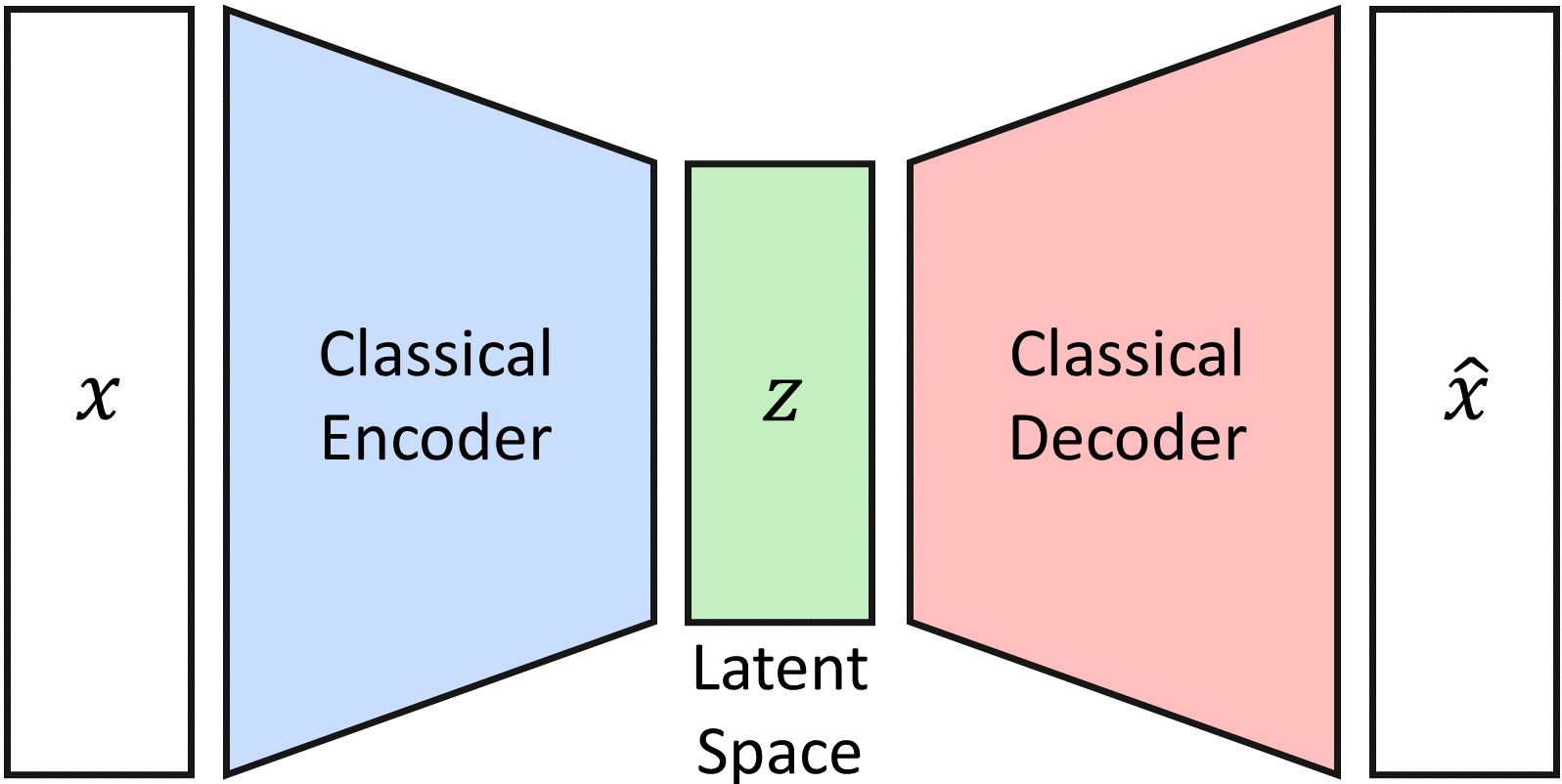}
    \caption{Architecture of classical convolutional autoencoder: The core design of a convolutional autoencoder consists of classical encoder and decoder components, each featuring crucial convolutional layers. These layers play a pivotal role in transforming the input image data $x$ into a reconstructed representation $\hat{X}$ through the latent space $Z$.}
    \label{fig:ccae}
\end{figure}

An autoencoder (AE) is an unsupervised NN comprising an encoder $E$ and a decoder $D$, coined in the 1980s~\cite{baldi2012autoencoders}. It reconstructs input image data within its hidden layers, using the input as the expected output. The CCAE enhances this by substituting fully connected layers with convolutional layers, optimizing it for spatial data tasks, and ensuring superior feature extraction and representation learning. Like a typical AE, the CCAE maintains input and output layer sizes but transforms the decoder with transposed convolutional layers for enhanced decoding capabilities~\cite{yan2023hybrid}. Fig.~\ref{fig:ccae} illustrates the architecture of the CCAE, showcasing the process of deconstructing and reconstructing input image data. Through this transformation, the encoded data traverses the latent space $Z$, enabling the model to learn and focus specifically on crucial components of the input image data.

Given an input image data $x$, the process of CCAE is defined as follows:
\begin{equation}
    \hat{x} = D(Z), Z = E(x),
\end{equation}
\revise{the most commonly used training loss for autoencoders is the L2 reconstruction loss, defined as:
}
\begin{equation}
    L_{AE}=\|x - \hat{x}\|^{2},
\end{equation}
and $\| \cdot \|^{2}$ denotes the L2-norm.

\section{The proposed method}\label{sec:method}
In this section, we present a comprehensive end-to-end methodology outlining the proposed approach. We first extensively review the proposed technique in its quantum form, after which we meticulously present the quantum circuit using the QAOA method. Finally, we explain how we fine tune and improve the quantum circuit through a training optimization process, focusing on a PSR technique.

\subsection{Quantum convolutional autoencoder (QCAE)}
\begin{figure*}[ht]
    \centering
    \includegraphics[width=\textwidth]{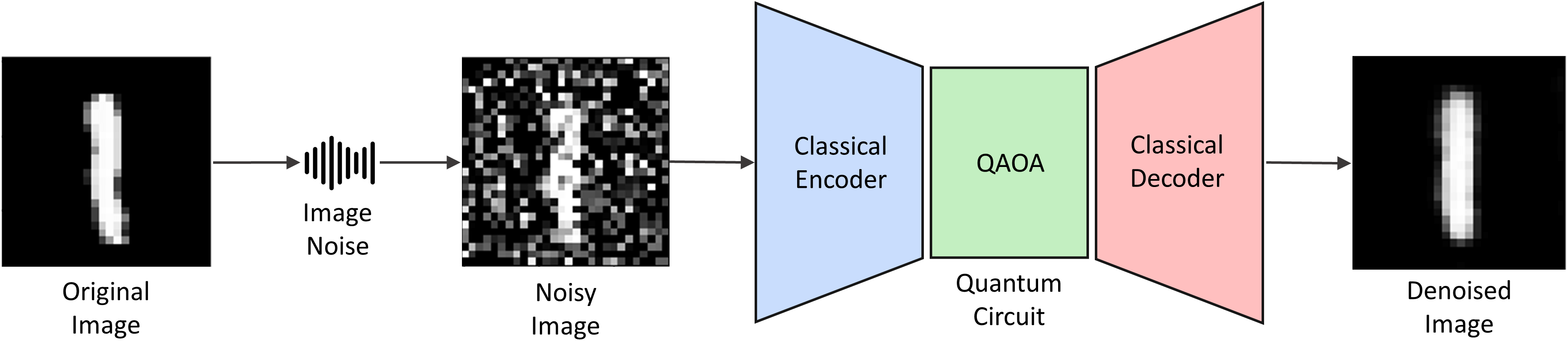}
    \caption{Architecture of quantum convolutional autoencoder, where the classical latent space is replaced with a QAOA circuit to efficiently denoise images.}
    \label{fig:qcae}
\end{figure*}

In this study, we introduce the QCAE architecture designed for image denoising. Leveraging the CCAE model as its foundation, the QCAE incorporates a QAOA circuit to replace the CCAE’s latent space. The QCAE process involves several key stages: \revise{i) pre-processing the input image data, which includes normalizing pixel values to the [0, 1] range,} ii) introducing \revise{Gaussian} noise to the input image data to generate a noisy input, \revise{iii) transforming the noisy input through classical convolutional layers to produce a latent representation, which is then encoded into a quantum state, iv) using a QAOA circuit to manipulate the quantum-encoded latent vector,} v) measuring the output of the QAOA circuit, and vi) generating denoised input image data through a classical decoder based on the QAOA circuit measurement results~\cite{hassan2024quantum}. Fig.~\ref{fig:qcae} showcases the QCAE architecture and the steps involved in image denoising. Given a random Gaussian noise-generating function $S$, the process involves taking an input data $x$ and subjecting it to the influence of $S$ to produce a noisy input image data version represented as $S(x)$. Subsequently, this noisy input $S(x)$ serves as the input for the QCAE. 

\newrevise{The training process of the QAOA circuit begins with input image data $x$ and a set of rotation gates at level $p$. First, the input image is encoded by a classical encoder $\vec{x} = E(x)$, where $\vec{x}$ represents the encoded latent vector~\cite{bravo2021quantum, shiba2019convolution}. This latent representation is further compressed into a linear form $y$ with dimension $2^p$, where $p \geq 1$ corresponds to the QAOA circuit depth. \newrevise{While this step reduces the dimensionality classically, the QAOA circuit operates in a quantum state space of dimension $2^n$, with $n$ denoting the number of qubits, thereby enabling access to richer high-dimensional representations}~\cite{rivas2021hybrid, kim2023classical}. Classical normalization and optimization methods are then applied to fine-tune the parameters and minimize the cost function. Fig.~\ref{fig:qaoa_input} illustrates this workflow, showing how the classical encoder prepares input data and optimizes it for QAOA circuits at various depths $p$.
}
% In the QCAE, the convolutional encoder is responsible for extracting and compressing crucial features from the input image data. Through a carefully designed sequence of convolutional layers, pooling layers, and activation functions, the encoder systematically reduces the dimensionality of the input image data. This process results in the creation of a condensed, low-dimensional latent representation that encapsulates the essential features of the input image~\cite{bravo2021quantum, shiba2019convolution}. \revise{While this is true for classical autoencoders, our quantum latent space, though implemented with a small number of qubits, operates in a much larger Hilbert space, enabling richer representational capacity.} By replacing the classical \revise{low-dimensional} latent space with the QAOA circuit, \revise{one can obtain representations that leverage quantum superposition and entanglement to preserve essential features in a higher-dimensional Hilbert space}~\cite{rivas2021hybrid, kim2023classical}. In contrast to the convolutional encoder, the convolutional decoder consists of a series of transposed convolutional layers, upsampling layers, and activation functions. Transposed convolutional layers essentially perform the inverse operation of convolutional layers. These layers augment the spatial dimensions of the feature maps, simultaneously acquiring the ability to reconstruct lost details of the encoded input image data.

\subsection{Quantum approximate optimization algorithm (QAOA)}
\begin{figure}[ht]
    \centering
    \includegraphics[width=0.75\textwidth]{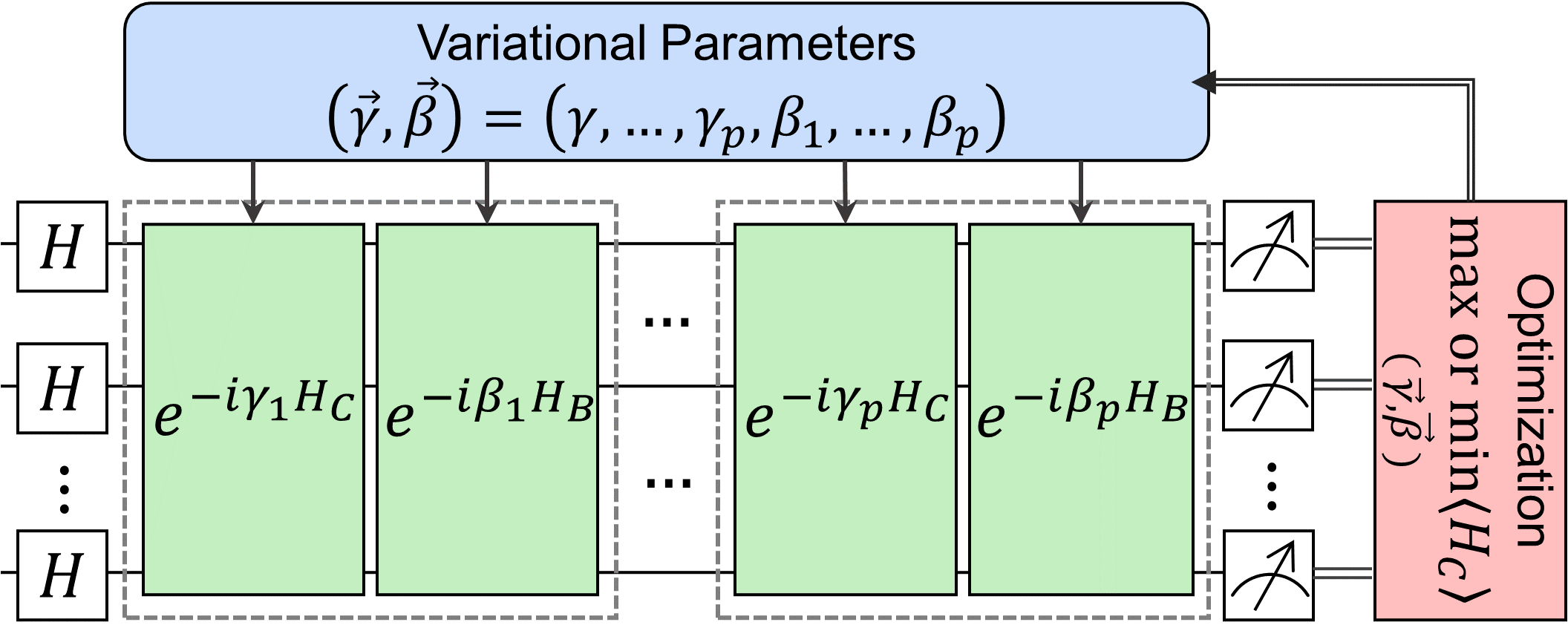}
    \caption{Schematic depicting the quantum approximate optimization algorithm, which integrates a quantum circuit and classical optimization to iteratively optimize variational parameters for enhanced performance.}
    \label{fig:qaoa}
\end{figure}

QAOA, originally proposed by Farhi et al.~\cite{farhi2014quantum}, is a heuristic approach specifically designed to tackle the maxcut problem. It belongs to the group of hybrid algorithms and requires, in addition to executing shallow quantum circuits, a classical optimization process to optimize the parameters and improve the quantum circuit itself~\cite{guerreschi2019qaoa, choi2019tutorial}.
\revise{The performance of QAOA continually improves with increasing values of p, where p denotes the number of alternating layers (depth) in the QAOA circuit, which controls the expressiveness of the variational ansatz}~\cite{farhi2016quantum}. As depicted in Fig.~\ref{fig:qaoa}, the maximum or minimum value of the objective function can be obtained by iteratively finding the optimal values for the parameters $\gamma$ and $\beta$. The iterative process for determining the optimal parameters relies on classical optimization methods. In the context of QCAE, QAOA is adapted to create a quantum latent space representation. This latent space aims to generate a compressed encoding of input image data, capturing essential patterns and relationships while reducing dimensionality. The QAOA circuit in QCAE leverages its optimization capabilities to discern quantum information from the input state, creating an optimal latent representation. This approach utilizes QAOA's problem-solving framework to efficiently encode input data in a quantum latent space, potentially offering advantages in data compression and feature extraction for image processing tasks. 

The algorithm is designed to identify a $n$-bit binary strings $z$, the objective function $f$ is defined as follows:
\begin{equation}
    f(z) : \{0, 1\}^n.
\end{equation}
We can map $f$ to the Hamiltonian $H_C$ as follows:
\begin{equation}
    H_C\vert z \rangle = f(z)\vert z \rangle.
\end{equation}
In which the phase operators are defined as follows:
\begin{equation}
    U_C(\gamma) = e^{-i \gamma H_C},
\end{equation}
where $\gamma$ is a parameter. The mixing Hamiltonian $H_B$ is defined as follows:
\begin{equation}
    H_B = \sum_{j=1}^{n}\sigma_{j}^{x},
\end{equation}
where $\sigma_{j}^{x}$ is the Pauli-X operator.  In addition, the mixing operators are defined as follows:
\begin{equation}
    U_B(\beta) = e^{-i \beta H_B},
\end{equation}
where $\beta$ is a parameter. We can define the state of the $p$-level QAOA by applying the phase operator and the mixing operator, defined as follows:
\begin{equation}
    \vert \vec{\gamma}, \vec{\beta} \rangle = U_C(\gamma_p)U_B(\beta_p) \cdots  U_C(\gamma_1)U_B(\beta_1)\vert s \rangle , 
\end{equation}
with an integer $p \geq 1$ and $s$ is the initial state, defined according to the superposition principle as follows:
\begin{equation}
    \vert s \rangle = \vert + \rangle^{\otimes n} = \frac{1}{\sqrt{2^n}}\sum_{z}\vert z \rangle.
\end{equation}
The expectation value of $H_C$ can be obtained as follows:
\begin{equation}
    \langle H_C \rangle := \langle \vec{\gamma}, \vec{\beta} \vert H_C \vert \vec{\gamma}, \vec{\beta} \rangle = \langle f \rangle_{(\vec{\gamma}, \vec{\beta})},
\end{equation}
where $f$ is the expectation value of the objective function.

\begin{figure}[ht]
    \centering
    \fbox{\includegraphics[width=0.75\textwidth]{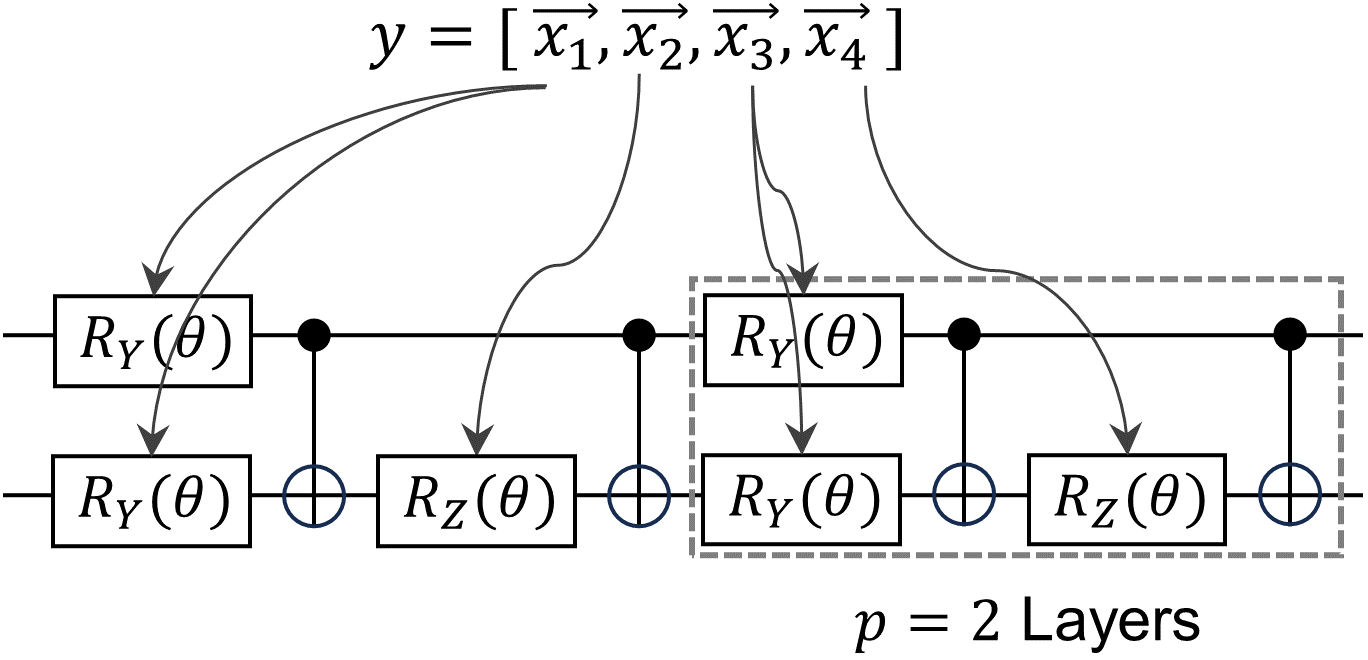}}
    \caption{Parameterizing the rotation gates of QAOA circuit with the optimized value of the encoded input image data.}
    \label{fig:qaoa_input}
\end{figure}

Here, we then introduce the process of injecting parameter values into rotation gates $R$ parameter placeholders. The training process of the QAOA circuit begins with input image data $x$ and a set of rotation gates $R$ at level $p$. First, the input image data are encoded using a classical encoder $\vec{x} = E(x)$, where $\vec{x}$ is an encoded input image data. Then, this data undergoes additional classical processing through convolutional layers, pooling layers, and is transformed into a linear form of $y$ to the size of $2p$, where $p \geq 1$ is the QAOA circuit level. \revise{While $y$ is classically compressed, the quantum circuit operates on a state space of dimension $2^n$, where $n$ is the number of qubits, providing access to high-dimensional quantum representations} Classical normalization and optimization techniques are applied to the variable $y$ to fine-tune the parameters, aiming to minimize the objective function (cost function). Fig.~\ref{fig:qaoa_input} illustrates the schematic illustrates the classical encoder in encoding and optimizing input image data for the QAOA circuit at any given $p$ levels.

\subsection{Parameter-shift rule optimization}
In this section, we present a PSR optimization method inspired by~\cite{wang2022qoc}. The PSR is a promising approach for evaluating the gradients of QAOA circuits on real quantum machines~\cite{crooks2019gradients, Wierichs_2022}. Gradients derived from these quantum circuits significantly contribute to the optimization phase of VQAs. In the context of hybrid quantum-classical algorithms, we first construct a quantum circuit and subsequently adjust its parameters to minimize an objective function or cost function~\cite{romero2017quantum}. In our proposed method, we compute the gradient of each parameter in the QAOA circuit by per the parameter and determining the difference between two forming a double shift, \textit{positive shift} and \textit{negative shift}, on outputs, without modifying the QAOA circuit structure or using any ancilla qubits~\cite{wang2022qoc}. Suppose we have an $n$-qubit QAOA circuit which is parameterized by $m$ parameters, 
$\theta = [\theta_1, \ldots, \theta_i, \ldots, \theta_m]$, the objective function $f(\theta)$ of a quantum circuit can be represented by a circuit function as follows:
\begin{equation}
    f(\theta) = \langle \psi \vert U(\theta_i)^{\dagger} Q U(\theta_i) \vert \psi \rangle,
\end{equation}
where $\theta_i$ is the scalar parameter whose gradient is to be computed and $U(\theta_i)$ is the unitary gate with a parameter $\theta_i$, which is the rotation gate in this paper. For notation simplicity, we will absorb the unitaries before $U(\theta_i)$ into $\langle \psi \vert$ and $\vert \psi \rangle$. Unitaries after $U(\theta_i)$ and observables are fused into $Q$. The parameterized unitary gates can be basically written in the following form:
\begin{equation}
    U(\theta_i) = e^{-\frac{i}{2}\theta_i H}, 
\end{equation}
where $H$ is the Hermitian generator of $U(\theta_i)$ with only two unique eigenvalues $+1$ and $-1$. In this way, the differences in the gradients of the circuit function $f$ with respect to $\theta_i$ are:
% \begin{equation}
%     \begin{array}{l}
%          \frac{\partial f(\theta)}{\partial \theta_i} = \frac{1}{2} \Bigl(f(\theta_{+}) - f(\theta_{-})\Bigl), \\ \\
%          \theta_{+} = [\theta_1, \dotsm, \theta_i + \frac{\pi}{2}, \dotsm, \theta_m], \\ \\
%          \theta_{-} = [\theta_1, \dotsm, \theta_i - \frac{\pi}{2}, \dotsm, \theta_m],
%     \end{array}
%     \label{eq:shifting}
% \end{equation}
\begin{equation}
\begin{array}{l}
\displaystyle
\frac{\partial f(\theta)}{\partial \theta_i}
= \frac{1}{2}\!\left(f(\theta_{+}) - f(\theta_{-})\right),\\[8pt]
\theta_{+} = [\,\theta_1,\, \ldots,\, \theta_i + \frac{\pi}{2},\, \ldots,\, \theta_m\,],\\[6pt]
\theta_{-} = [\,\theta_1,\, \ldots,\, \theta_i - \frac{\pi}{2},\, \ldots,\, \theta_m\,],
\end{array}
\label{eq:shifting}
\end{equation}

where $\theta_{+}$ and $\theta_{-}$ are the \textit{positive shift} and \textit{negative shift} of $\theta$, respectively. This method precisely computes the gradient with respect to $\theta_i$, avoiding any approximation errors. We then apply the softmax function to the objective function obtained from measurements $f(\theta)$ as the predicted probability distribution for \textit{positive shift} and \textit{negative shift}. Subsequently, we compute the cross entropy between the predicted probability distribution $pd$ and the target distributed $td$ as the classification loss $L$ as follows:
\begin{equation}
    L(\theta) = -td^{T} \cdot softmax(f(\theta)) = -\sum_{j=1}^{n}td_j\log{pd_j},
\end{equation}
where
\begin{equation}
    pd_j = \frac{e^{f_j(\theta)}}{\sum_{j=1}^{n}e^{f_j(\theta)}}
\end{equation}
Then the gradient of the loss function with respect to $\theta_i$ can be defined as follows:
\begin{equation}
    \frac{\partial L(\theta)}{d\theta_i} = \Bigl(\frac{\partial L(\theta)}{\partial f(\theta)}\Bigl)^T \times \; \frac{\partial f(\theta)}{\partial \theta_i}
    \label{eq:gradient}
\end{equation}
Here $\frac{\partial f(\theta)}{\partial \theta_i}$ can be computed on a QAOA circuit utilizing the PSR, while $\frac{\partial L(\theta)}{\partial f(\theta)}$ can be efficiently calculated on classical devices through backpropagation, facilitated by differentiation frameworks, e.g., PyTorch~\cite{paszke2019pytorch} and TensorFlow~\cite{abadi2016tensorflow}.
\begin{figure}[ht]
    \centering
    \includegraphics[width=0.75\textwidth]{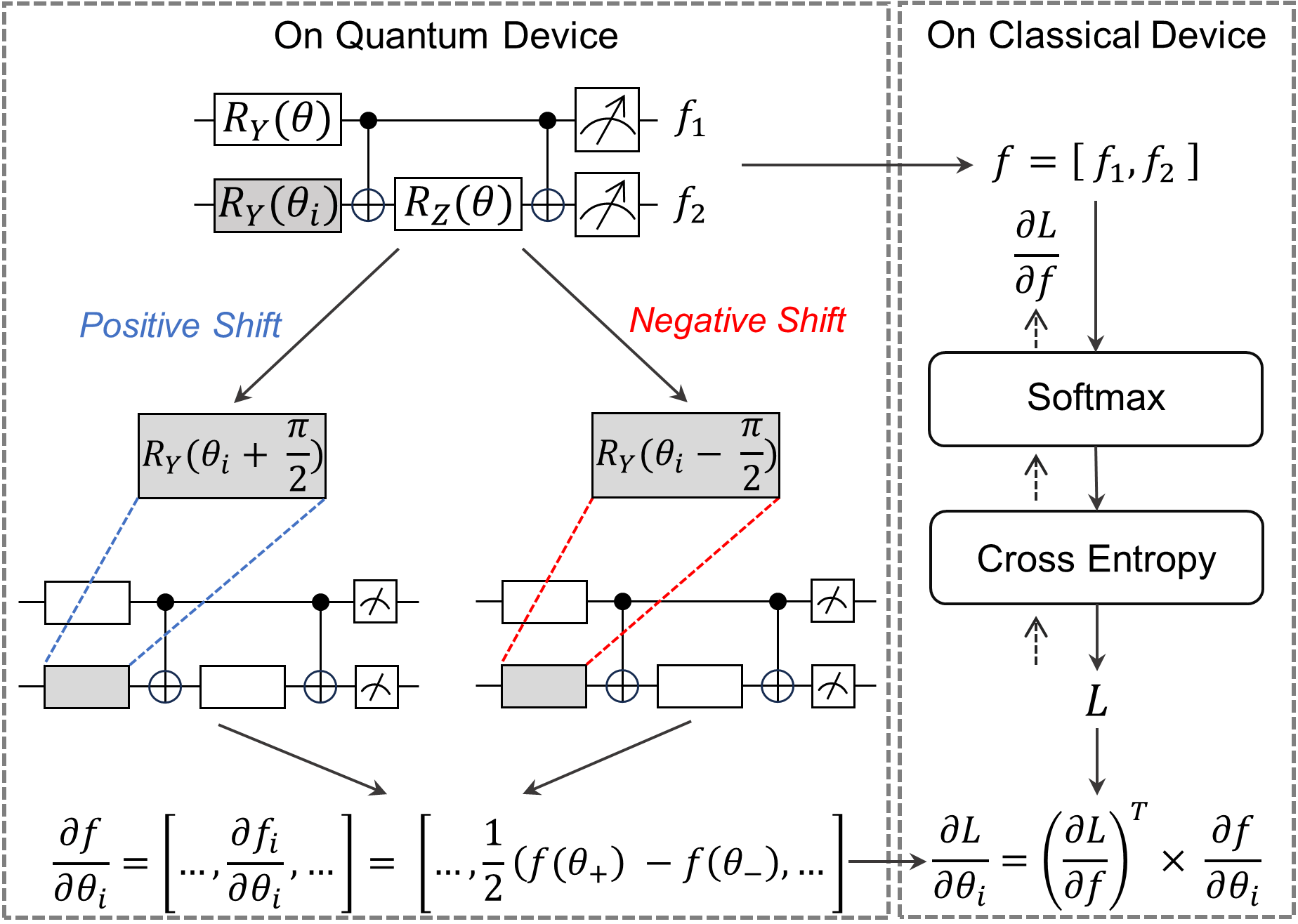}
    \caption{Efficient gradient computation through the technique PSR on a hybrid QAOA circuit.}
    \label{fig:parameter_shift_rules}
\end{figure}

Fig.~\ref{fig:parameter_shift_rules} visually depicts the procedural steps of the technique PSR. In each iteration, we perform dual shifts of the parameter $\theta_i$ by $+\frac{\pi}{2}$ and $-\frac{\pi}{2}$, respectively. After each parameter shift, we execute the shifted QAOA circuit on either a local quantum simulator or a quantum device. Upon obtaining the results from the two shifted circuits, we apply Equation~\ref{eq:shifting} to calculate the upstream gradient. Following the PSR, we then execute the QAOA circuit without any parameter shift and record the measurement results. Subsequently, we apply softmax and cross-entropy functions to the obtained logits, ultimately producing training loss. The backpropagation process is initiated solely from the loss to the logits, generating downstream gradients. Finally, the gradient is computed by taking the dot product between the upstream and downstream gradients, as outlined in Eq.~\ref{eq:gradient}, resulting in the final gradient.

\begin{figure}[t]
     \centering
     \begin{subfigure}[c]{0.23\textwidth}
         \centering
         \includegraphics[width=0.57\textwidth]{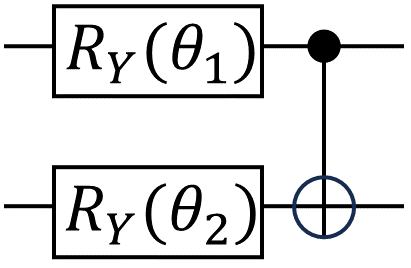}
         \caption{Circuit A}
         \label{fig:circuita}
     \end{subfigure}
     \hfill 
     \begin{subfigure}[c]{0.23\textwidth}
         \centering
         \includegraphics[width=0.67\textwidth]{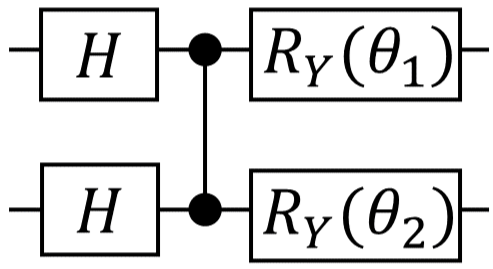}
         \caption{Circuit B}
         \label{fig:circuitb}
     \end{subfigure}
     \hfill \vspace{3mm}
     \begin{subfigure}[c]{0.23\textwidth}
         \centering
         \includegraphics[width=\textwidth]{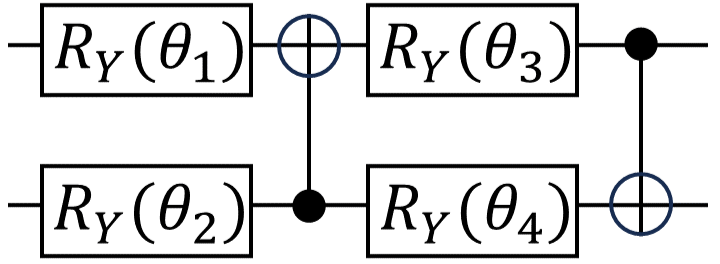}
         \caption{Circuit C}
         \label{fig:circuitc}
     \end{subfigure}
     \hfill
     \begin{subfigure}[c]{0.23\textwidth}
         \centering
         \includegraphics[width=0.95\textwidth]{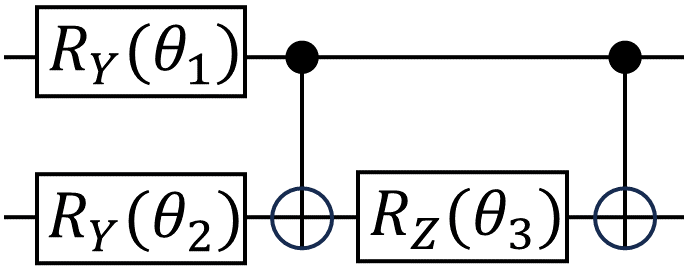}
         \caption{Ours}
         \label{fig:ours}
     \end{subfigure}
    \caption{PQCs utilized in the QAOA circuit. The PQCs (a), (b), and (c) are adapted from~\cite{kim2023classical, hur2022quantum}, whereas (d) is our designed QAOA circuit.}
    \label{fig:circuits}
\end{figure}

\section{Experimental setups}\label{sec:setup}
In this section, we outline the datasets used to validate our proposed method and describe the experimental environments for its implementation.

\subsection{Datasets}
We perform both CCAE and QCAE on the well-known MNIST dataset. This dataset comprises handwritten digits presented as grayscale images. The MNIST dataset comprises 60,000 samples in the training set and 10,000 samples in the testing set, each represented by pixels arranged in a shape of $28 \times 28 \times 1$. To incorporate \revise{Gaussian} noise into the datasets, we used a function that adds random Gaussian noise \revise{with standard deviation values $\sigma = \{0.25, 0.5, 0.75, 1\}$} to the images. The original and noisy images are visualized in Fig.~\ref{fig:qcae}.
\newrevise{Although restricting MNIST to digits 0 and 1 yields 12,000 training and 2,000 test samples, in this study we used smaller subsets to reduce computational cost and accommodate quantum hardware limitations. Specifically, we selected the first 2,000 samples of digits 0 and 1 from the training set (total 4,000 training images) and the first 200 samples of digits 0 and 1 from the test set (total 400 test images). Gaussian noise was applied only to the training set during model optimization, while the test set remained unseen throughout training and was used exclusively for evaluation. This ensures that no overlap occurs between training and testing data.}
% \revise{For the MNIST dataset, we selected two specific classes (digits 0 and 1) to reduce computational complexity and qubit requirements during training and evaluation on quantum simulators and real quantum hardware.} For a fair comparison between CCAE and QCAE, the training datasets comprise the initial 2,000 samples, \revise{while a subset of the MNIST test split was used for evaluation.
A separate validation set was not used during training due to the dataset size; model performance was assessed directly on the test set after training. \revise{While MNIST is widely used for benchmarking, it has limitations in demonstrating the generalizability and real-world applicability of machine learning models, particularly in the quantum domain, as outlined in~\cite{bowles2024better}.}

\subsection{Environments}
The experiments were conducted using the PyTorch~\cite{paszke2019pytorch} and Qiskit~\cite{qiskit} frameworks. The proposed QCAE is tested on both a local quantum simulator equipped with a quantum assembly (QASM) simulation and a 27-qubit IBMQ Mumbai quantum machine. For noisy simulators, we leverage the \revise{quantum} noise model from IBM Mumbai and inject it into the QASM simulator. For notation simplicity, we designate the QASM simulator as the \textit{noiseless} simulator, the noisy QASM simulator as the \textit{noisy} simulator, and the actual quantum machine as the \textit{real} machine.

For both the CCAE and QCAE, the encoder and decoder architectures comprise three 2D convolutional and 2D transposed convolutional layers, respectively. The only difference lies in the latent space, which is substituted with a QAOA circuit in the case of QCAE. In the CCAE model, the latent space is established through a fully connected layer network, with the classical encoder and decoder employing the LeakyReLU activation function. The training spanned 50 epochs using the Adam optimizer~\cite{kingma2014adam} and the mean squared error (MSE) as the loss function. Due to cost and time constraints, we conducted the training in both \textit{noiseless} and \textit{noisy} environments, as it required many iterations. To validate the real-world applicability, the testing phase was conducted in a \textit{real} environment, ensuring the proposed model's reliability in practical scenarios.

We meticulously execute and perform experiments that encompass a comprehensive array of QAOA circuits. These circuits incorporate diverse configurations of single-qubit and two-qubit gate operations. Our circuit designs were inspired by prior research~\cite{kim2023classical, hur2022quantum}, and their visual representation is illustrated in Fig.~\ref{fig:circuits}.

\begin{figure}[ht]
    \centering
    \includegraphics[width=0.75\textwidth]{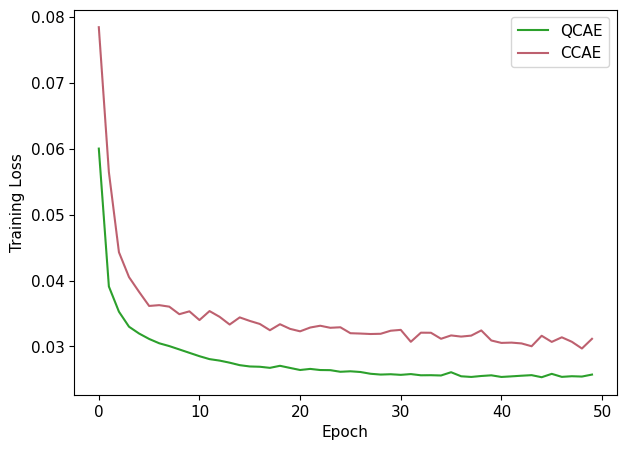}
    \caption{\revise{Training loss comparison between CCAE and QCAE over 50 epochs.}}
    \label{fig:training_ssim}
\end{figure}

\section{Experimental results}\label{sec:results}
In this section, we evaluate the performance of our proposed QCAE model on the denoising MNIST dataset. First, we compare the training loss and evaluation of QCAE with CCAE. Second, we analyze the denoising performance
of QCAE in \textit{noiseless}, \textit{noisy}, and \textit{real} scenarios. Finally, we conduct ablation studies by varying QAOA circuits and layers to enhance understanding and optimize its overall performance.

\subsection{Training of the CCAE and QCAE}
\begin{figure}[ht]
    \centering
    \includegraphics[width=0.48\textwidth]{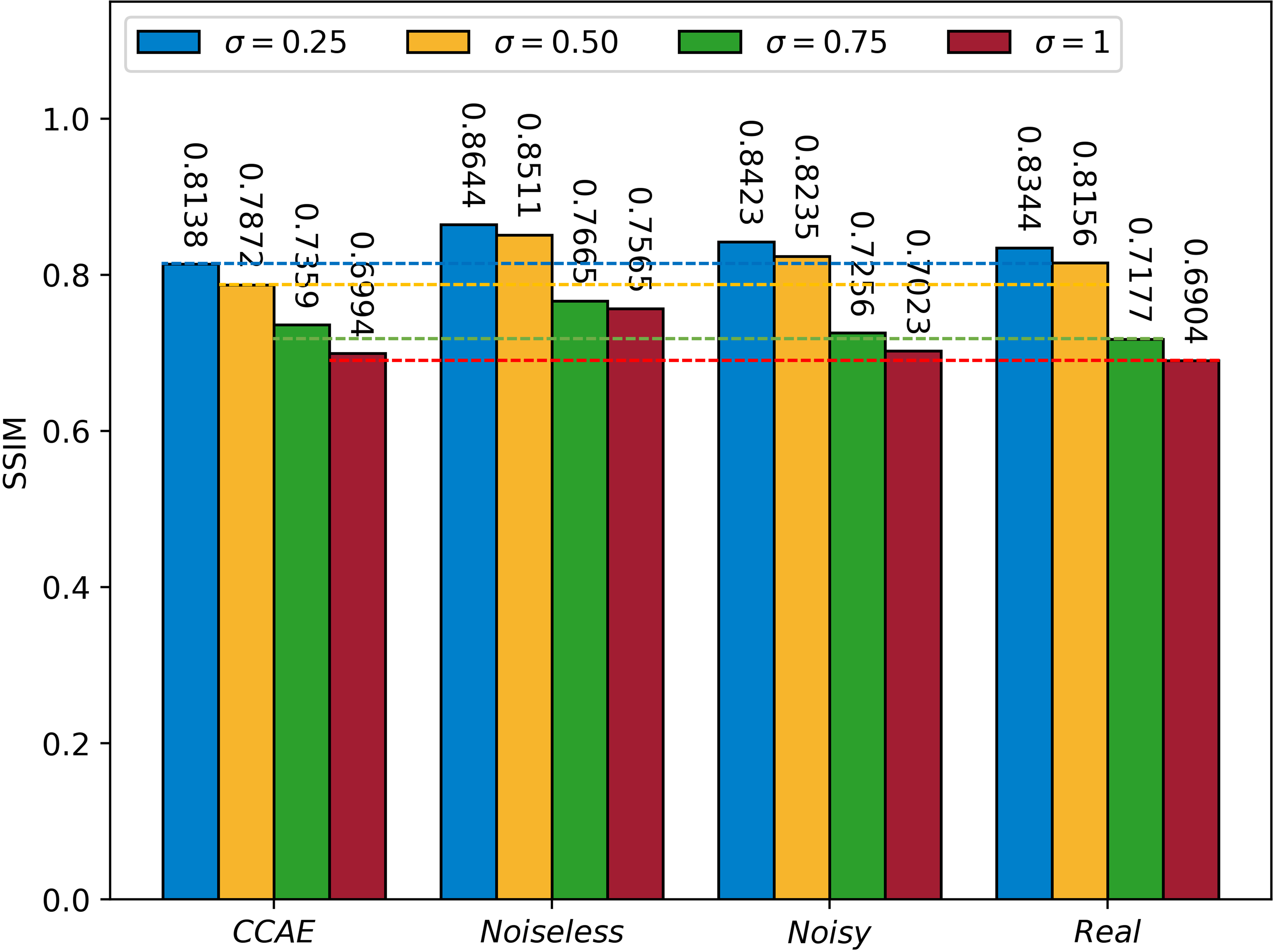}
    \caption{Comparative analysis of the CCAE and specifically QCAE performance under varying conditions, including \textit{noiseless}, \textit{noisy}, and \textit{real} scenarios, with noise factor variations represented by $\sigma = \{0.25, 0.5, 0.75, 1\}$. \revise{Evaluation was conducted on the MNIST test set (digits 0 and 1 only).}}
    \label{fig:performance}
\end{figure}

In this evaluation, two binary classes were extracted from the MNIST dataset (0 and 1). To gain insights into the distinctions between denoised samples generated by both the CCAE and QCAE models and the original samples, we employ a metric called the structural similarity index measure (SSIM) for analyzing image quality~\cite{tanchenko2014visual}, ranging from 0 (worst) to 1 (best).
\newrevise{Fig.~\ref{fig:training_ssim} illustrates the training loss over 50 epochs, where QCAE consistently outperforms CCAE. The corresponding SSIM values are presented in Fig.~\ref{fig:performance} across all scenarios (\textit{Noiseless}, \textit{Noisy}, and \textit{Real}). QCAE achieves higher SSIM values than CCAE in most cases. However, at higher noise levels ($\sigma = 0.75$ and $1$) in the Real and Noisy scenarios, its SSIM values are slightly lower than those of CCAE, reflecting the impact of quantum hardware noise.}
% To illustrate the training performance, the left side of Fig.~\ref{fig:training_ssim} displays the training loss. Our proposed QCAE consistently outperforms CCAE, achieving a lower training loss. The right side of Fig.~\ref{fig:training_ssim} illustrates the SSIM values over 50 training epochs, demonstrating that QCAE generates denoised images closely resembling and showcasing higher quality compared to those produced by CCAE. This distinction is particularly evident when considering the average of SSIM 0.45 for CCAE and SSIM 0.75 for QCAE.
Overall, these findings highlight that QCAE consistently delivers robust results when employed for denoising images compared with CCAE.

\begin{figure}[ht]
    \centering
    \includegraphics[width=0.482\textwidth]{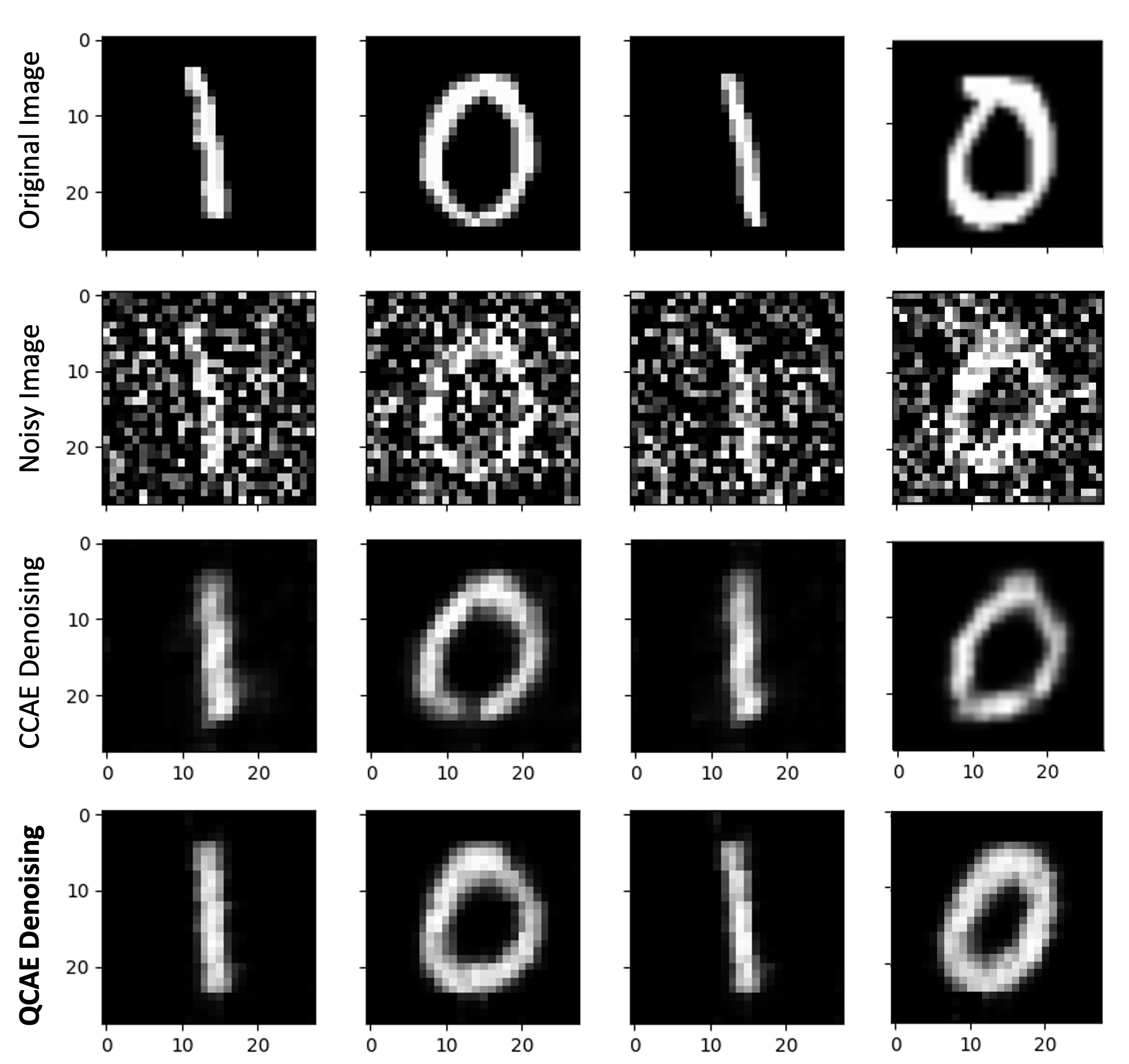}
    \caption{Denoising results on MNIST datasets. The top row displays original images, and the second row shows their noisy counterparts. The third row showcases images denoised with CCAE, and the last row features images denoised using the proposed QCAE method. \revise{All images are from the MNIST test set.}}
    \label{fig:denoising}
\end{figure}

Fig.~\ref{fig:denoising} depicts original images alongside those with added \revise{Gaussian} noise, where high corruption levels render them barely visible to the human eye. The denoising results generated using both QCAE and CCAE models are shown alongside the noisy and original images. \newrevise{Notably, in the third row and second column, the CCAE struggles to fully denoise or reconstruct the image, highlighting a limitation}. Furthermore, CCAE faces challenges in effectively denoising, resulting in denoised results that exhibit a subtle blurring effect when compared to the sharper outcomes achieved by QCAE, with an average SSIM for CCAE at 0.45 and the SSIM for QCAE at 0.75. This performance gap emphasizes the effectiveness of our proposed QCAE, which outperforms CCAE as its classical counterpart.

\subsection{Denoising performance}
Here, we conducted denoising of the MNIST images across three distinct scenarios: \textit{noiseless}, \textit{noisy}, and \textit{real}. To effectively demonstrate the performance of the QCAE, we introduced variations in Gaussian noise factors applied to the images because the MNIST does not carry noise. For each scenario, 
\revise{we tested 12 different configurations of Gaussian noise $\sigma = \{0.25, 0.5, 0.75, 1\}$ across three environments (noiseless, noisy, real), and selected the combination that yielded the highest SSIM score for further experiments.}
The QCAE performance is evaluated for the denoising capability in both quantum simulations and actual machines.

\begin{figure}[ht]
    \centering
    \includegraphics[width=0.47\textwidth]{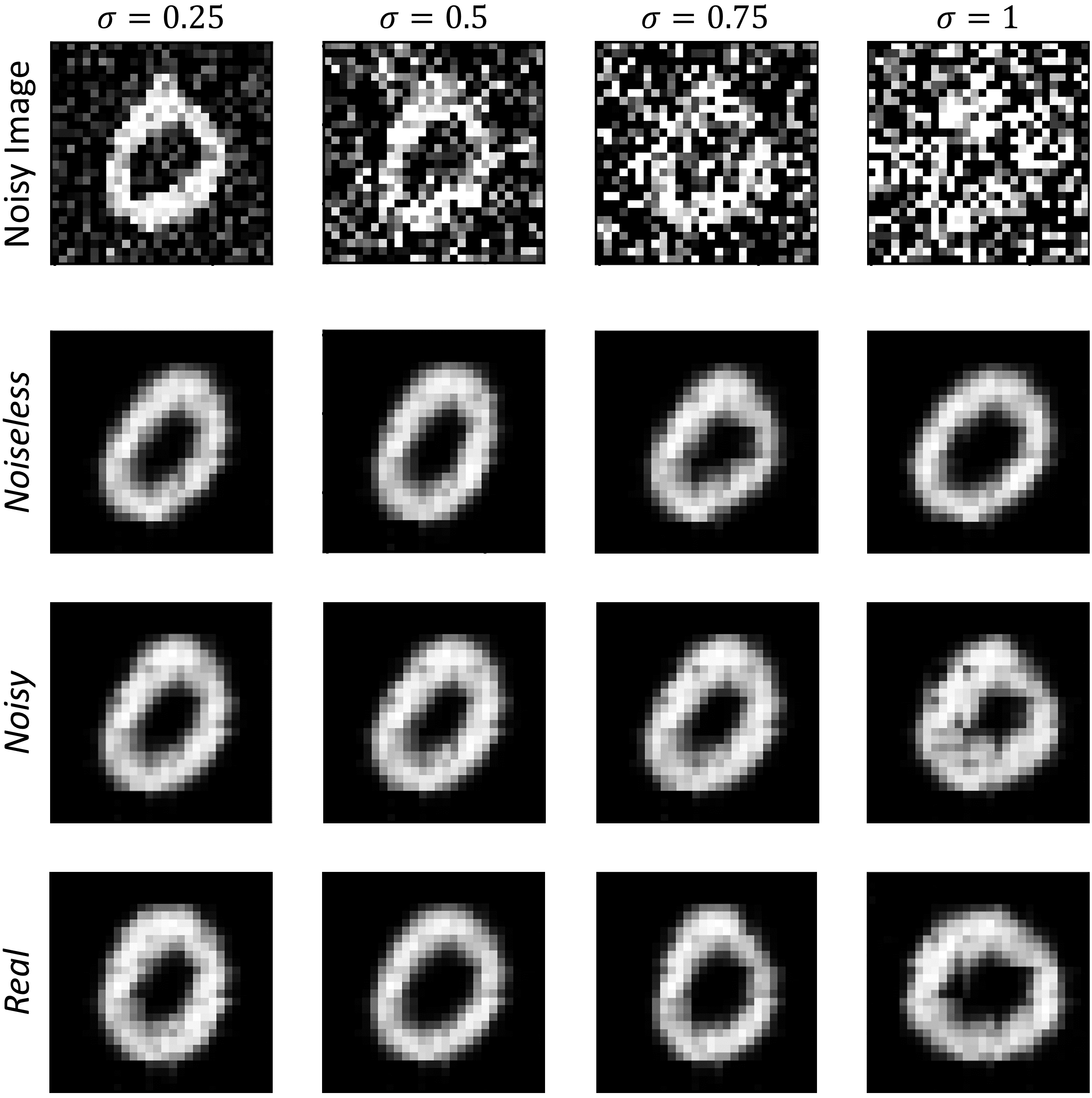}
    \caption{Performance evaluation of QCAE on noisy images across a range of Gaussian noise factors. The top row presents noisy images with $\sigma = \{0.25, 0.5, 0.75, 1\}$. The second row showcases denoised images in a \textit{noiseless} quantum simulation, whereas the third row showcases denoised images in a \textit{noisy} quantum simulation. The fourth row displays denoised images with a \textit{real} quantum machine.}
    \label{fig:performance_images}
\end{figure}

To analyze the extent of QCAE denoising capabilities, we utilize it to restructure, denoise, and generate images from a set of \revise{Gaussian} noisy inputs with varying noise factors values. Even in the presence of high levels of noise, characterized by a noise factor of $\sigma = 1$, images that appear indistinguishable to the human eye are effectively processed by QCAE. The algorithm consistently delivers accurate and visually appealing results, demonstrating its robust performance in handling noisy images. Despite testing in \textit{noisy} and \textit{real} quantum machine environments, the image maintains readability and closely resembles the original, although with a slight degradation compared to a \textit{noiseless} environment. 
% Denoising results, along with variations in Gaussian noise levels and quantum machine environments, are depicted in Fig.~\ref{fig:performance_images} \revise{are based on digit ‘0’ images selected from the same MNIST test subset (digits 0 and 1) used in prior evaluations.
\newrevise{As depicted in Fig.~\ref{fig:performance_images}, denoising results with variations in Gaussian noise levels and quantum machine environments are shown for digit ‘0’ images selected from the same MNIST test subset (digits 0 and 1) used in prior evaluations.}
This ensures consistency across all noise levels and quantum environments.

\subsection{Comparison of PQCs and PSR optimization}
\begin{figure}[hb]
    \centering
    \includegraphics[width=0.482\textwidth]{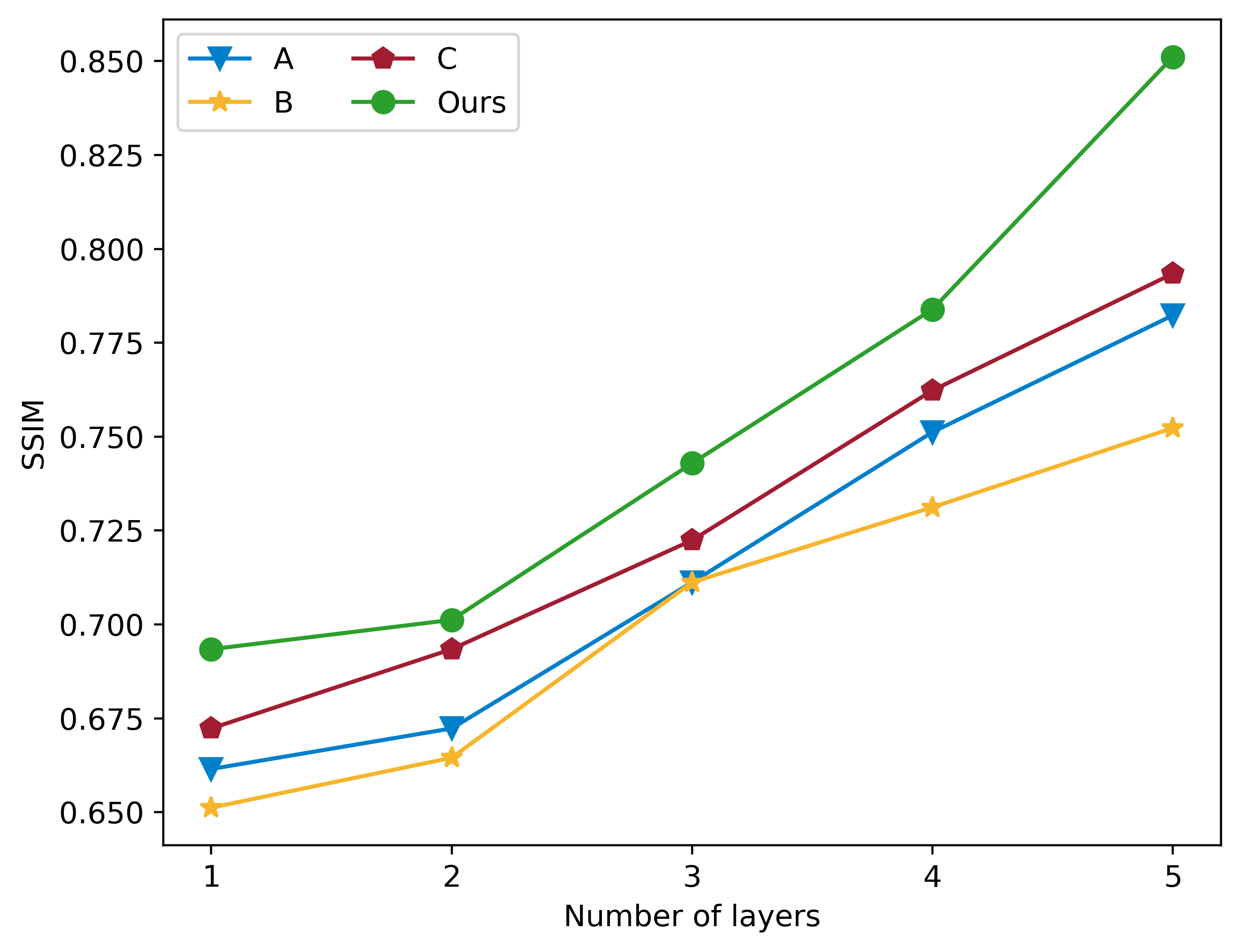}
    \caption{The SSIM value of $p$ ranges from 1 to 5 across all circuits, indicating that better solutions are associated with longer circuit layers. \revise{Evaluation was conducted on the same MNIST test subset (digits 0 and 1) used in earlier sections.}}
    \label{fig:ablation}
\end{figure}

In this study, we conducted experiments employing various quantum circuits sourced from~\cite{kim2023classical, hur2022quantum}, and comparing them to our proposed approach with and without PSR optimization. The experiments were executed with identical configurations, including the number of qubits, epochs, $\sigma = 0.5$, optimization using the PSR, a \textit{noiseless} environment, and the MNIST dataset. In addition, we varied the number of $p$ layers to investigate how circuit depth influences the obtained results. The $p$-value results were validated across a range from 1 to 5 and subsequently categorized into groups corresponding to the respective quantum circuits. Fig.~\ref{fig:ablation} presents the SSIM values obtained from the various circuit types depicted in Fig.~\ref{fig:circuits}. Compared to alternative circuits, our proposed approach showcases superior performance across these circuits. Specifically, our proposed QAOA circuit consistently outperforms other circuits, showcasing an average improvement of approximately 25\% in the SSIM value.

We also conducted comprehensive experiments on all circuits, both with and without PSR, resulting in a noteworthy reduction in SSIM values when tested without PSR, as shown in Table~\ref{tab:with_without_psr}. Remarkably, our proposed circuit demonstrated the ability to draw comparisons with other circuits, even in the absence of PSR. One plausible explanation is that $R_Z$ gates can be implemented with greater efficiency compared to other single-qubit rotation operations~\cite{mckay2017efficient}. As we observed, the circuits exhibit lower and significantly higher SSIM values with and without PSR, respectively. This underscores the significance of incorporating PSR optimization to enhance performance. It is noteworthy that the $p$ value significantly influences the results, underscoring its high impact on the outcomes. However, as the parameter $p$ increases, there is a corresponding growth in the circuit execution time. Additionally, our proposed method requires a minimum of two qubits for the efficient denoising of images. It also illustrates the progressive increase in SSIM values corresponding to the varying number of $p$ layers. As additional layers are added, the QAOA showcases the potential for enhanced approximations towards the optimal solution. Furthermore, it presents the results of experiments conducted on QAOA, both with and without PSR optimization. The findings emphasize the significance of employing this optimization technique to identify a lower-cost function, leading to higher accuracy. Through the utilization of the PSR optimization, there is a notable enhancement in SSIM value. 
% This enhancement showcases an improvement of up to 35\% when compared with non-optimized conditions.

\begin{table}
\caption{\label{tab:with_without_psr}The SSIM value for parameter $p$ varies between 1 and 5 across
all circuits, both with and without the utilization of PSR optimization techniques.}
\begin{indented}
\lineup
\item[] \begin{tabular}{cccccc}
\br
Circuits & p = 1  & p = 2  & p = 3  & p = 4  & p = 5  \\
\mr
\multicolumn{6}{c}{Without PSR optimization}          \\
\mr
A        & 0.4371 & 0.4398 & 0.4423 & 0.4480 & 0.4512 \\
B        & 0.4302 & 0.4357 & 0.4423 & 0.4521 & 0.4553 \\
C        & 0.4493 & 0.4523 & 0.4567 & 0.4623 & 0.4663 \\
Ours     & 0.6503 & 0.6618 & 0.7034 & 0.7235 & 0.7496 \\
\mr
\multicolumn{6}{c}{With PSR optimization}             \\
\mr
A        & 0.6615 & 0.6723 & 0.7112 & 0.7512 & 0.7823 \\
B        & 0.6512 & 0.6645 & 0.7112 & 0.7312 & 0.7523 \\
C        & 0.6723 & 0.6934 & 0.7224 & 0.7623 & 0.7934 \\
Ours     & 0.6934 & 0.7012 & 0.7429 & 0.7839 & 0.8511 \\
\br
\end{tabular}
\end{indented}
\end{table}

\section{Conclusion}\label{sec:conclusion}
This study introduced a QCAE model that leverages the power of QAOA to seamlessly substitute for the conventional AE, serving as the latent space. This substitution improves data learning in a higher-dimensional space. We demonstrated the efficiency of our proposed QCAE method through its application to MNIST image denoising.  The proposed QCAE method exhibited superior denoising performance, as it achieved lower training loss and increased SSIM, compared to the CCAE method—a finding validated by experimental results—demonstrating its superior capacity to preserve image fidelity. The developed model maintained its superior denoising capabilities when applied to the MNIST dataset. Furthermore, we examined QAOA performance to determine its effectiveness across diverse circuit configurations and layer variations. Finally, we conducted an experiment using PSR optimization to evaluate its impact on QAOA circuit accuracy. This investigation underscores the crucial role of optimization in achieving desirable outcomes. However, we remark that state-of-the-art and out-of-the-box classical algorithms cannot be easily outperformed by current quantum computers and algorithms.
Notably, we did not include the mixer QAOA typically used in the QAOA circuits. The reason for this omission is that our primary focus was on evaluating the effectiveness of the QAOA mixing Hamiltonian in enhancing the latent space representation within the QCAE model. In standard QAOA, the mixing Hamiltonian is indeed a fundamental component and is typically chosen as a transverse field. This Hamiltonian helps to explore the solution space by inducing transitions between different quantum states. By isolating the mixing Hamiltonian, we aimed to specially assess its impact on denoising performance without the added complexity introduced by the mixer QAOA.
The potential future research directions are as follows: First, we aim to evaluate the effectiveness of our proposed approach by conducting tests on real-world practical images and providing evidence to validate its viability. Second, we aim to investigate how different optimization methods affect the performance of our proposed model on the MNIST dataset. Furthermore, given that notice noise in real quantum machines can significantly impact the outcomes of the method, employing noise mitigation techniques is required to address and alleviate these effects. \revise{Future work will also explore how the dimensionality of the latent space influences the QCAE’s representational capacity and learning performance, which could provide insights for optimizing the balance between model complexity and denoising accuracy.}

\section*{Code available}
The code that supports the findings of this study is openly available in the Github repository: \href{https://github.com/tarakit/quantum-autoencoder-denoising}{Quantum Autoencoder}.

\section*{CRediT authorship contribution statement}

\textbf{Tara Kit:} Conceptualization, Methodology, Validation, Visualization, Software, Writing – review \& editing. \textbf{Kimsay Pov and Kimleang Kea:} Writing - review \& Validation. \textbf{Won-Du Chang:} Writing – review, Validation, Visualization. \textbf{Hee Chul Park:} Writing – review, Validation. \textbf{Youngsun Han:} Supervision, Validation, Visualization, Writing –
review \& editing.

\section*{Acknowledgments}
This research was supported by Quantum Computing based on Quantum Advantage challenge research (RS-2023-00257994) through the National Research Foundation of Korea (NRF) funded by the Korean government (MSIT).

\bibliographystyle{iopart-num}
\bibliography{manuscript}

\end{document}